\documentclass{jnmp01b}

\usepackage{amsmath}

\numberwithin{equation}{section}

\def\a{\alpha}
\def\b{\beta}
\def\d{\delta}
\def\g{\gamma}
\def\o{\omega}
\def\s{\sigma}
\def\vn{\varnothing}
\def\vp{\varphi}
\def\lra{\longrightarrow}
\newcommand{\fg}{{\mathfrak g}}
\newcommand{\R}{\mathbb R}
\newcommand{\ad}{{\mathop{\mbox{ad}}\nolimits}}
\newcommand{\Int}{{\mathop{\mbox{Int}}\nolimits}}
\newcommand{\fu}{{\mathfrak U}}
\newcommand{\fs}{{\mathfrak S}}
\newcommand{\G}{\mathbb G}
\newcommand{\ca}{{\mathcal A}}
\newcommand{\Ad}{{\mathop{\mbox{Ad}}\nolimits}}
\newcommand{\Ker}{{\mathop{\mbox{Ker}}\nolimits\,}}
\newcommand{\cg}{{\mathcal G}}
\newcommand{\cm}{{\mathcal M}}
\newcommand{\cb}{{\mathcal B}}
\newcommand{\ct}{{\mathcal T}}
\newcommand{\cn}{{\mathcal N}}
\newcommand{\frs}{{\mathfrak s}}
\newcommand{\U}{{\mathcal U}}

\begin{document}

\renewcommand{\evenhead}{T.A. Ivanova and A.D. Popov}
\renewcommand{\oddhead}{On Symmetries of Chern-Simons and BF Topological Theories}


\thispagestyle{empty}

\begin{flushleft}
\footnotesize \sf
Journal of Nonlinear Mathematical Physics \qquad 2000, V.7, N~4,
\pageref{firstpage}--\pageref{lastpage}.
\hfill {\sc Article}
\end{flushleft}

\vspace{-5mm}

\copyrightnote{2000}{T.A. Ivanova and A.D. Popov}

\Name{On Symmetries of Chern-Simons and BF Topological Theories}

\label{firstpage}

\Author{T.A. IVANOVA and A.D. POPOV}

\Adress{Bogoliubov Laboratory of Theoretical Physics\\
JINR, 141980 Dubna, Moscow Region, Russia\\
E-mail: ita@thsun1.jinr.ru and popov@thsun1.jinr.ru}

\Date{Received March 6, 2000; Revised June 20, 2000;
Accepted July 11, 2000}

\begin{abstract}
\noindent
We describe constructing solutions of the field equations of
Chern-Simons and topological BF theories in terms of deformation
theory of locally constant (flat) bundles. Maps of flat connections
into one another (dressing transformations) are considered.
A method of calculating (nonlocal) dressing symmetries in
Chern-Simons and topological BF theories is formulated.
\end{abstract}


\section{Introduction}

Let $X$ be an oriented smooth manifold of dimension $n$, $G$ a
semisimple Lie group, $\fg$ its Lie algebra, $P$ a principal
$G$-bundle over $X$, $A$ a connection 1-form on $P$ and
$F_A=dA+A\wedge A$ its curvature. We assume that $X$ is a
compact manifold without boundary.

A connection 1-form $A$ on $P$ is called {\it flat} if its curvature
$F_A$ vanishes,
\begin{equation*}
dA+A\wedge A=0.
\tag{1.1a}
\end{equation*}
Locally eqs.(1.1a) are solved trivially, and on any sufficiently small
open set $U\subset X$ we have $A=-(d\psi )\psi^{-1}$, where $\psi (x)\in G$.
So, locally $A$ is a pure gauge. But globally eqs.(1.1a) are nontrivial,
and finding their solutions is not an easy problem.

It is well known that in the case $n=dim_{\R}X=3$ eqs.(1.1a) are the field
equations of Chern-Simons theories which describe flat
connections on $G$-bundles over 3-manifolds $X$. Quantum observables
of these theories are topological invariants of $X$~\cite{Sc,Wi}.

{}For a generalization of Chern-Simons theories to arbitrary dimensions,
in addition to a connection $A$ on a $G$-bundle $P\to X$, one considers
a $(n-2)$-form $B$ with values in the adjoint  bundle $\ad P=P\times_G\fg$.
Using the fields
$B$ and $F_A$, one introduces so-called BF theories~\cite{Ho,BT},
for which the variation of the action w.r.t. $B$ gives eqs.(1.1a), and the
variation of the action w.r.t. $A$ gives the equations
\begin{equation*}
d_AB:=dB+A\wedge B-B\wedge A=0,
\tag{1.1b}
\end{equation*}
where $d_A$ is the covariant differential on $P$. For more details and
references see~\cite{BBRT}.

The purpose of our paper is to describe the symmetry group of
eqs.(1.1) arising as equations of motion in Chern-Simons
and topological BF theories. Under the symmetry group of a system
of differential equations we understand the group that maps solutions
of this system into one another. Only gauge (and related to them)
symmetries  of eqs.(1.1) forming a small subgroup in the symmetry
group have been considered in the literature.

In this paper we show how {\it all} symmetries of eqs.(1.1) can be
calculated with the help of {\it deformation theory} methods. In fact,
finding solutions and symmetries of eqs.(1.1) will be
reduced to solving functional equations on some matrices. To illustrate
the method, we write down some nontrivial explicit solutions of these
functional equations and describe symmetries corresponding to them.

\section{Definitions and notation}

In this section we recall some definitions to be used in
the following and fix the notation.
In Sect.2.1 we introduce bundles $\Int P$ of Lie groups, bundles
$\ad P$ of Lie algebras and sheaves of their sections~\cite{Hir, GrH}.
In considering sheaves of non-Abelian groups we follow~\cite{Hir} and
the papers~\cite{F, D, On}. In Sect.2.2 we recall
definitions~\cite{Hir, F, D, On} of cohomology sets of
manifolds with values in the sheaves of Lie groups.

\subsection{Some bundles and sheaves}

Let $\{P_i\}$ be a set of representatives of topological
equivalence classes of $G$-bundles over $X$. We fix $P\in\{P_i\}$
and choose a {\it good} covering $\fu =\{U_\a , \a\in I\}$
of the manifold $X$, which is always possible. For such a covering
$\fu$ each nonempty finite intersection $U_{\a_1}\cap ...\cap U_{\a_q}$
is diffeomorphic  to an open ball in $\R^n$. Let $f=\{f_{\a\b}\}$,
$\ f_{\a\b}: U_\a\cap U_\b\to G$ be transition functions determining the
bundle $P$ in a fixed trivialization.

Recall that a map of a subset $U\subset X$ into some set is called
{\it locally constant} if it is constant on each connected component
of the set $U$. The fibre bundle $P$ is called {\it locally constant}
if its transition functions $\{f_{\a\b}\}$ are locally constant.
All bundles associated with such a bundle are also called locally constant.
In the following, we shall consider deformations of the locally constant
bundle $P$. They are described in terms of cohomologies of the manifold
$X$ with values in sheaves of sections of some bundles associated with $P$.
Below we introduce these sheaves.

Let us consider the bundle of groups $\Int P=P\times_GG$
($G$ acts on itself by internal automorphisms:
$h\mapsto ghg^{-1}, \ h,g\in G$) associated with $P$.
Consider the sheaf $\fs_P$ of {\it smooth} sections of
the bundle $\Int P$ and its subsheaf $\G_P$ of {\it locally
constant} sections. A section $\psi$ of the sheaf  $\G_P$ ($\fs_P$)
over an open set $U\subset X$ is described by a collection $\{\psi_\a\}$
of locally constant (smooth) $G$-valued functions $\psi_\a$ on
$U\cap U_\a\ne\vn$ such that $\psi_\a = f_{\a\b}\psi_\b f_{\a\b}^{-1}$
on  $U\cap U_\a\cap U_\b\ne\vn$.

We also consider the adjoint bundle $\ad P=P\times_G\fg$ of Lie
algebras and denote by $\ca^q_P$ the sheaf of smooth $q$-forms
on $X$ with values in the bundle $\ad P$ ($q=1,2,...$). The space of
sections of the sheaf $\ca^q_P$ over an open set $U$ is the space
$\ca^q_P(U)\equiv\Gamma (U, \ca^q_P)$ of smooth $\fg$-valued $q$-forms
$\ca^{(\a )}$ on $U\cap U_\a\ne\vn$ such that $\ca^{(\a )}=f_{\a\b}
\ca^{(\b )}f_{\a\b}^{-1}$ on $U\cap U_\a\cap U_\b\ne\vn$. In particular,
for global sections of the sheaves $\ca^q_P$ over $X$
one often uses the notation
\begin{equation*}
\Omega^q(X, \ad P):= \ca_P^q(X)\equiv \Gamma (X, \ca_P^q).
\tag{2.1}
\end{equation*}

The sheaf $\fs_P$ (and $\G_P$) acts on the sheaves $\ca^q_P, q = 1,2,...$,
with the help of the adjoint representation. In particular, for any open
set $U\subset X$ we have
\begin{gather*}
A\mapsto \Ad_\psi A=\psi A\psi^{-1} + \psi d\psi^{-1} ,
\tag{2.2a}
\\
{}F\mapsto \Ad_\psi F=\psi F\psi^{-1}  ,
\tag{2.2b}
\\
B\mapsto \Ad_\psi B=\psi B\psi^{-1},
\tag{2.2c}
\end{gather*}
where $\psi \in \fs_P (U)$, $A\in \ca^1_P(U)$, $F\in \ca^2_P(U)$, $B\in
\ca^{n-2}_P(U)$. Of course,  $\psi d\psi^{-1}=0$ if $\psi\in\G_P(U)$.

Denote by $i : \G_P\to\fs_P$ an embedding of $\G_P$ into
$\fs_P$. We define a map $\d^0 : \fs_P \to \ca^1_P$ given for any open
set $U$ of the space $X$ by the formula
\begin{equation*}
\d^0\psi := -(d\psi )\psi^{-1},
\tag{2.3}
\end{equation*}
where $\psi \in \fs_P (U)$, $\d^0 \psi \in \ca^1_P(U)$. Let us also
introduce an operator $\d^1 : \ca_P^1 \to \ca_P^2$, defined for an open
 set $U\subset X$ by the formula
\begin{equation*}
\d^1A := dA + A\wedge A ,
\tag{2.4}
\end{equation*}
where $A\in \ca^1_P(U)$, $\d^1A\in \ca^2_P(U)$. In other words, the maps
of sheaves $\d^0 : \fs_P\to \ca^1_P$ and $\d^1 : \ca^1_P\to\ca^2_P$ are
defined by means of localization. Finally, we denote by $\ca_P$ the
subsheaf in $\ca^1_P$, consisting of locally defined $\fg$-valued
1-forms $A$ such that $\d^1A = 0$, i.e. sections $A\in \ca_P (U)$
of the sheaf $\ca_P =\Ker \d^1$ over $U\subset X$ satisfy eqs.(1.1a).

\subsection{\v{C}ech cohomology for sheaves of non-Abelian groups}

Let $S$ be a sheaf coinciding with either the sheaf $\G_P$ or the sheaf
$\fs_P$ introduced in Sect.2.1. We consider a good open covering
$\fu =\{U_\a\}$ of $X$ and sections of the sheaf $S$ over $U_{\a_0}\cap ...
\cap U_{\a_q}\ne\vn$. In the trivialization over $U_{\a_k}$, we can
represent $h\in\Gamma (U_{\a_0}\cap ...\cap U_{\a_m}\cap ...\cap U_{\a_k}
\cap ...\cap U_{\a_q}, S)$ by matrices $h^{(\a_k)}_{\a_0...\a_m...\a_k...
\a_q}$, and in the trivialization over $U_{\a_m}$, $h$ is represented by
\begin{equation*}
h^{(\a_m)}_{\a_0...\a_m...\a_k...\a_q} = f_{\a_m\a_k}
h^{(\a_k)}_{\a_0...\a_m...\a_k...\a_q}f_{\a_m\a_k}^{-1},
\tag{2.5a}
\end{equation*}
where $\{f_{\a_m\a_k}\}$ are the transition functions of the bundle $P$.
If we do not write a superscript of sections of bundles, we mean that
\begin{equation*}
h_{\a_0...\a_q}:=h_{\a_0...\a_q}^{(\a_0)}.
\tag{2.5b}
\end{equation*}

A {\it $q$-cochain} of the covering $\fu$ with values in $S$ is a collection
$h=\{h_{\a_0...\a_q}\}$ of sections of the sheaf $S$ over nonempty intersections
$U_{\a_0}\cap ...\cap U_{\a_q}$. A set of $q$-cochains is denoted by
$C^q(\fu , S)$; it is a {\it group} under the pointwise multiplication. In
particular, for 0-cochains $h=\{h_\a\}\in C^0(\fu , S)$ with the
coefficients in $S$ we have $h_\a\in\Gamma (U_\a , S)$ and for 1-cochains
$h=\{h_{\a\b}\}\in C^1(\fu , S)$ we have $h_{\a\b}\in \Gamma
(U_\a\cap U_\b , S)$.

Subsets of {\it cocycles} $Z^q (\fu , S)\subset C^q(\fu , S)$ for
$q=0,1$ are defined as follows:
\begin{gather*}
Z^0(\fu , S):=\{h\in C^0(\fu , S): h_\a f_{\a\b}h_\b^{-1}=f_{\a\b}\
\mbox{on}\  U_\a\cap U_\b\ne\vn\},
\tag{2.6}
\\[1ex]
Z^1(\fu , S):=\{h\in C^1(\fu , S): h_{\a\b}f_{\a\b}h_{\b\g}
f_{\b\g}h_{\g\a}f_{\g\a}=1\
\mbox{on}\  U_\a\cap U_\b\cap U_\g\ne\vn\}.
\tag{2.7a}
\end{gather*}
It follows from (2.7a) that
\begin{gather*}
h_{\a\b}^{(\a )}h_{\b\a}^{(\a )}=1
\quad\mbox{on}\quad U_{\a}\cap U_{\b}\ne\vn,
\tag{2.7b}
\\
h_{\a\a}=1\quad\mbox{on}\quad  U_{\a},
\tag{2.7c}
\end{gather*}
where $h_{\b\a}^{(\a )} =f_{\a\b}h_{\b\a}f_{\a\b}^{-1}$ (see (2.5)).
We see from (2.6) that $Z^0(\fu , S)$ coincides with the group
\begin{equation*}
H^0(X,S):=\Gamma (X,S)
\tag{2.8}
\end{equation*}
of global sections of the sheaf $S$. The set $Z^1(\fu , S)$ is not a
group if the structure group $G$ of the bundle $P$ is a non-Abelian
group.

{}For $h\in C^0(\fu ,S),\ \phi\in Z^1(\fu ,S)$ we define the left action
$T_0$ of the group $C^0(\fu ,S)$ on the set $Z^1(\fu ,S)$ by the
formula
\begin{equation*}
T_0(h,\phi )_{\a\b}=h_\a\phi_{\a\b}(h_{\b}^{(\a )})^{-1}\equiv
h_\a\phi_{\a\b}f_{\a\b}h^{-1}_\b f^{-1}_{\a\b}.
\tag{2.9a}
\end{equation*}
A set of orbits of the group $C^0(\fu ,S)$ in $Z^1(\fu ,S)$ is called
a {\it 1-cohomology set} and denoted by $H^1(\fu ,S)$. In other words,
two cocycles $\phi ,\tilde\phi\in Z^1(\fu ,S)$  are called
{\it equivalent} (cohomologous) if
\begin{equation*}
\tilde\phi_{\a\b}=h_\a\phi_{\a\b}(h_{\b}^{(\a )})^{-1}
\tag{2.9b}
\end{equation*}
for some $h\in C^0(\fu ,S)$, and the 1-cohomology set
\begin{equation*}
H^1(\fu ,S):=C^0(\fu ,S)\backslash Z^1(\fu ,S)
\tag{2.10}
\end{equation*}
is a set of equivalence classes of 1-cocycles. Since the chosen
covering $\fu =\{U_\a\}$ is {acyclic} for the sheaf $S$,
we have  $H^1(\fu ,S)=H^1(X ,S)$. For more details see
~\cite{Hir, GrH, F, D, On}.

\section{Moduli spaces}

Chern-Simons and topological BF theories give a field-theoretic
description of locally constant (flat) bundles which is discussed
in this section. In Sect.3.1 we introduce the de Rham 1-cohomology
set describing the moduli space of flat connections and give the
de Rham description of moduli of covariantly constant fields $B$.
Using homological algebra methods~\cite{F, D, On}, in Sect.3.2 we
present the equivalent \v{C}ech description of the moduli spaces of flat
connections and covariantly constant fields $B$.

\subsection{Flat connections, $\ad P$-valued forms and de Rham cohomology}

We denote by  $\cg_P$ the infinite-dimensional group of gauge
transformations, i.e. smooth automorphisms of the bundle $P\to X$
which induce the identity transformation of $X$. Put another way,
$\cg_P$ is the group of global smooth sections of the bundle $\Int P$
\cite{AHS}, and therefore we have
\begin{equation*}
\cg_P = H^0(X,\fs_P)\equiv \Gamma (X,\fs_P)=Z^0(\fu ,\fs_P)
\tag{3.1}
\end{equation*}
with the notation of Sect.2.

The space of all connections on $P$ is an infinite-dimensional affine
space modelled on the vector space $H^0(X,\ca^1_P)=\Omega^1(X, \ad
P)$ of global sections of the sheaf $\ca^1_P$ and therefore the space
of flat connections on $P$ can be identified with the space
$H^0(X,\ca_P)$ of global sections of the sheaf $\ca_P =\Ker \d^1$ (see
(2.4)). The group
(3.1) acts on the space $H^0(X,\ca_P)$ on the right by the formula
\begin{equation*}
A\mapsto \Ad_{g^{-1}} A=g^{-1} A g + g^{-1} dg ,
\end{equation*}
where $g\in \cg_P$,
 and we
introduce the de Rham 1-cohomology set $H^1_{d_A;P}(X)$ as a set
of orbits of the non-Abelian group $H^0(X,\fs_P)$ in the set
$H^0(X,\ca_P)$,
\begin{equation*}
H^1_{d_A;P}(X):=H^0(X,\ca_P)/H^0(X, \fs_P)=\{A\in\Omega^1(X,\ad P):
dA+A\wedge A=0\}/\cg_P .
\tag{3.2}
\end{equation*}
This definition is a generalization to $\ad P$-valued 1-forms $A$ and
the covariant exterior derivative $d_A$ of the standard definition of
the 1st de Rham cohomology group
\begin{equation*}
H^1_{d}(X)=\frac{\{\o\in\Omega^1(X): d\o=0\}}
{\{\o=d\vp, \vp\in\Omega^0(X)\}},
\tag{3.3}
\end{equation*}
where $d$ is the exterior derivative.

\medskip

\noindent
{\bf Remark.} \ For an Abelian group $G$, the group (3.1) of gauge
transformations and the set $H^0(X, \ca_P)$ are Abelian groups.
In this case, (3.2) is reduced in fact to the definition (3.3) of the
standard de Rham cohomology.

\medskip

The definition (3.2) is nothing but the definition of the moduli
space $\cm_P$ of flat connections on $P$,
\begin{equation*}
\cm_P = H^1_{d_A;P}(X)=H^0(X,\ca_P)/H^0(X, \fs_P).
\tag{3.4}
\end{equation*}
So, the moduli space $\cm_P$ of flat connections on $P$ is the space
of gauge inequivalent solutions to eqs.(1.1a) as it should be.

Equations (1.1b) are linear in $B$. For any fixed flat
connection $A$, the moduli space of solutions to eqs.(1.1b) is a vector
space
\begin{equation*}
\cb_A=H^{n-2}_{d_A;P}(X)=\frac{\{B\in\Omega^{n-2}(X, \ad P): d_AB=0\}}
{\{B=d_A\Phi , \Phi\in\Omega^{n-3}(X, \ad P)\}},
\tag{3.5}
\end{equation*}
since solutions of the form $B=d_A\Phi$ are considered as trivial
(topological ``symmetry'' $B\mapsto B+d_A\Phi$)~\cite{BBRT}.

Nontrivial solutions of eqs.(1.1b) form the vector space
$\cb_A$ depending on a solution $A$ of eqs.(1.1a). The space of solutions
to eqs.(1.1) forms a vector bundle $\ct_P\to H^0(X,\ca_P)$, the base
space of which is the space  $H^0(X,\ca_P)$ of solutions to eqs.(1.1a),
and fibres of the bundle at points $A\in H^0(X,\ca_P)$ are the
vector spaces $\cb_A$ of nontrivial solutions to eqs.(1.1b).

Notice that the gauge group $\cg_P=H^0(X,\fs_P)$ acts on solutions
$B$ of eqs.(1.1b) by the formula
\begin{equation*}
B\mapsto \Ad_{g^{-1}}B=g^{-1}Bg,
\tag{3.6}
\end{equation*}
where $g\in \cg_P$. Therefore, identifying points $(A,B)\in\ct_P$
and $(g^{-1}Ag + g^{-1}dg, g^{-1}Bg)\in\ct_P$ for any $g\in\cg_P$,
we obtain the moduli space
\begin{equation*}
\cn_P=\ct_P/\cg_P
\tag{3.7}
\end{equation*}
of solutions to eqs.(1.1). The space $\cn_P$ is the vector bundle over the
moduli space $\cm_P$ of flat connections with fibres at points
$[A]\in\cm_P$ isomorphic to the vector space $\cb_A$. We denote by $[A]$
the gauge equivalence class of a flat connection $A$.

\subsection{Flat connections, $\ad P$-valued forms and \v{C}ech cohomology}

By using definitions of the sheaves $\G_P, \fs_P$ and $\ca_P$, it is not
difficult to verify that the sequence of sheaves
\begin{equation*}
{e}\lra\G_P\stackrel{i}\lra \fs_P\stackrel{\d^0}\lra
\ca_P \stackrel{\d^1}\lra  e
\tag{3.8}
\end{equation*}
is exact. Here $e$ is a marked element of the considered sets (the identity
in the sheaf $\G_P\subset \fs_P$ and zero in the sheaf $\ca_P$). From
(3.8) we obtain the exact sequence of cohomology
sets~\cite{F, D, On}
\begin{equation*}
e\lra H^0(X,\G_P)\stackrel{i_*}\lra H^0(X,\fs_P)\stackrel{\d^0_*}\lra
H^0(X,\ca_P) \stackrel{\d^1_*}\lra H^1(X,\G_P)
\stackrel{\rho}\lra H^1(X,\fs_P),
\tag{3.9}
\end{equation*}
where the map $\rho$
coincides with the canonical embedding induced by the embedding of sheaves
$i: \G_P\to \fs_P $.

By definition the set of equivalence classes of locally constant
$G$-bundles  over $X$ is parametrized by the \v{C}ech 1-cohomology
set $H^1(X,\G_{P_0})\simeq H^1(X,\G_{P})$~\cite{Hir,F,D}.
The kernel $\Ker\rho =\rho^{-1}(e)$ of the map $\rho$ coincides
with a subset of those elements $[\hat f f^{-1}]$ from $H^1(X,\G_P)$
which are mapped
into the class $e\in H^1(X, \fs_P)$ of bundles $\hat P$ diffeomorphic
to the bundle $P$ defined by the 1-cocycle $f$. In terms of transition
functions $\{f_{\a\b}\}$ in $P$ and $\{\hat f_{\a\b}\}$ in $\hat P$
this means that there exists a smooth 0-cochain $\psi =\{\psi_\a\}\in
C^0(\fu , \fs_P)$ such that
\begin{equation*}
\hat f_{\a\b}=\psi^{-1}_\a f_{\a\b}\psi_\b ,
\tag{3.10}
\end{equation*}
i.e. cocycles $f$ and $\hat f$ are smoothly cohomologous.

By virtue of the exactness of the sequence (3.9), the space $\Ker\rho$
is bijective to the quotient space (3.2).
So we have
\begin{equation*}
\cm_P= H^1_{d_A;P}(X)\simeq \Ker\rho\subset H^1(X,\G_P),
\tag{3.11}
\end{equation*}
i.e. there is a one-to-one correspondence between the moduli space $\cm_P$
of flat connections on $P$ and the moduli space $\Ker\rho\subset H^1(X,\G_P)$
of those locally constant bundles $\hat P$ which are diffeomorphic to the
bundle $P$.   Notice that to $\hat f=\{\psi^{-1}_\a f_{\a\b}\psi_\b\}$,
representing $[\hat f f^{-1}]\in\Ker\rho$ with fixed $f$, there corresponds
a flat connection
$A=\{A^{(\a )}\}=\{\psi_\a d\psi^{-1}_\a\}\in H^0(X, \ca_P)$ representing
$[A]\in H^1_{d_A;P}(X)$.

\medskip

\noindent
{\bf Remark.} \ It is well known that the bundle $P\to X$ is always
diffeomorphic to the trivial bundle $X\times G$ if $X$ is a paracompact
and simply connected manifold~\cite{Hir}. It is also known that if the
structure group $G$ is connected and simply connected, and $dim_{\R}X\le 3$,
then all $G$-bundles over $X$ are topologically trivial, i.e.
$H^1(X,\fs_P)=e$. In both the cases we have
\begin{equation*}
H^1_{d_A;P}(X)\simeq H^1(X,\G_P),
\tag{3.12}
\end{equation*}
since now $\Ker\rho =H^1(X,\G_P)$. The bijections (3.11) and (3.12)
are non-Abelian variants of the isomorphism between \v{C}ech and de Rham
cohomologies.

\medskip

The moduli space (3.5) of solutions to eqs.(1.1b) can also be described
in terms of \v{C}ech cohomology. Namely, notice that for a fixed flat
connection $A=\{\psi_\a d\psi^{-1}_\a\}$ a general solution of eqs.(1.1b)
has the form
\begin{equation*}
B=\psi B_0\psi^{-1}=\{\psi_\a B^{(\a )}_0 \psi^{-1}_\a\},
\tag{3.13a}
\end{equation*}
where $\psi =\{\psi_\a\}\in C^0(\fu ,\fs_P)$, and $B_0$ is an $\ad
\hat P$-valued $(n-2)$-form satisfying the equation
\begin{equation*}
dB_0=0. \tag{3.13b}
\end{equation*}
Therefore the space $H^{n-2}_{d_A;P}(X)$ of nontrivial solutions
of eqs.(1.1b) is isomorphic to the standard $(n-2)$th de Rham cohomology
group
\begin{equation*}
H^{n-2}_{d;\hat P}(X)=\frac{\{B_0\in\Omega^{n-2}(X,\ad\hat P): dB_0=0\}}
{\{B_0=d\Phi_0,\ \Phi_0\in\Omega^{n-3}(X, \ad\hat P)\}}
\tag{3.14}
\end{equation*}
for forms with values in the bundle $\ad\hat P$.

{}Formulae (3.13) mean that the space $H^{n-2}_{d_A;P}(X)$ is the
``dressed'' space $H^{n-2}_{d;\hat P}(X)$, and we have
\begin{equation*}
H^{n-2}_{d_A;P}(X) \simeq H^{n-2}_{d;\hat P}(X).
\tag{3.15a}
\end{equation*}
On the other hand, it is well known that
\begin{equation*}
H^{n-2}_{d;\hat P}(X)\simeq H^{n-2}(X, {\bf g}_{\hat P}),
\tag{3.15b}
\end{equation*}
where $H^{n-2}(X, {\bf g}_{\hat P})$ is the $(n-2)$th \v{C}ech cohomology
group of the manifold $X$ with the coefficients in the sheaf
${\bf g}_{\hat P}$ of locally constant sections of the bundle $\ad \hat P$.

\section{Construction of solutions}

In this section we discuss constructing solutions of eqs.(1.1)
in some more detail. Namely, in Sect.4.1 we consider a collection
of group-valued smooth functions defining maps between smooth
and locally constant trivializations of flat bundles. These functions
para\-me\-trize flat connections and satisfy first order differential
equations. The solution space of these equations is discussed in Sect.4.1.
In the \v{C}ech approach flat bundles are described by locally constant
transition functions. In Sect.4.2 we discuss
functional matrix equations on transition functions and the space
of their solutions. In Sect.4.3 we describe the moduli space of flat
connections as a double coset space and discuss the correspondence
between the \v{C}ech and de Rham descriptions of flat connections.

\subsection{Differential compatibility equations}

As before, we fix a good covering $\fu =\{U_\a\}$ of $X$, trivializations
of the bundle $P$ over $U_\a$'s and locally constant transition
functions $\{f_{\a\b}\}$ on $U_\a\cap U_\b\ne\vn$, $df_{\a\b}=0$, $\a ,
\b\in I$. Any solution of equations (1.1a) on an open set $U_\a$ (a
{\it local} solution) has the form
\begin{equation*}
A^{(\a )}=-(d\psi_\a )\psi^{-1}_\a ,
\tag{4.1a}
\end{equation*}
where $\psi_\a (x)$ is a smooth $G$-valued function on $U_\a$.
If we find  solutions (4.1a) for all $\a\in I$, we obtain a collection
$\psi =\{\psi_\a\}\in C^0(\fu ,\fs_P)$. If $U_\a\cap U_\b\ne\vn$,
then the solutions (4.1a) on $U_\a$ and $U_\b$ are not independent and
must satisfy the compatibility conditions
\begin{equation*}
A^{(\a )}=f_{\a\b}A^{(\b )}f_{\a\b}^{-1}
\tag{4.1b}
\end{equation*}
on $U_\a\cap U_\b$. From eqs.(4.1) the {\it differential} compatibility
equations
\begin{equation*}
(d\psi_\a )\psi_\a^{-1}= f_{\a\b} (d\psi_\b )\psi_\b^{-1}f_{\a\b}^{-1}
\tag{4.2}
\end{equation*}
follow. Since we are looking for {\it global} solutions $A=\{A^{(\a)}\}$
of eqs.(1.1a), eqs.(4.2) must be satisfied for any $\a ,\b\in I$ such
that $U_\a\cap U_\b\ne\vn$.

Any global solution of eqs.(1.1b) has the form
\begin{equation*}
B=\psi B_0\psi^{-1} =\{\psi_\a B_0^{(\a )}\psi_\a^{-1}\}=\{B^{(\a )}\},
\tag{4.3a}
\end{equation*}
where $\psi =\{\psi_\a\}$ satisfy eqs.(4.2), and $B_0$ is an arbitrary
(global) solution of eqs.(3.13b). The compatibility conditions for
$B^{(\a )}$ and $B^{(\b )}$ on $U_\a\cap U_\b\ne\vn$ have the form
\begin{equation*}
B^{(\a )}=f_{\a\b}B^{(\b )} f_{\a\b}^{-1}.
\tag{4.3b}
\end{equation*}
After substituting (4.3a) into (4.3b), we obtain that
\begin{equation*}
B^{(\a )}_0=\hat f_{\a\b}B^{(\b )}_0 \hat f_{\a\b}^{-1},
\tag{4.3c}
\end{equation*}
where $\hat f_{\a\b}=\psi_\a^{-1} f_{\a\b}\psi_\b ,\ d\hat f_{\a\b}=0$.
Thus, one can easily construct solutions of eqs.(1.1b) if one knows
solutions of eqs.(1.1a) or (4.2). That is why in the following we shall
concentrate on describing solutions of eqs.(1.1a) and (4.2).

Differential equations (4.2) are equations on 0-cochains $\psi =\{\psi_\a\}
\in C^0(\fu , \fs_P)$. We denote by $C^0_f(\fu , \fs_P)$ the space of
solutions $\psi =\{\psi_\a\}$ of eqs.(4.2), $C^0_f(\fu , \fs_P)\subset
C^0(\fu , \fs_P)$. Recall that according to the definitions of Sections
2 and 3, the solution space of eqs.(1.1a) is $Z^0(\fu ,\ca_P)
\equiv H^0 (X,\ca_P)$. Restricting the map $\d^0$ to $C^0_f(\fu , \fs_P)$,
we obtain the map
\begin{equation*}
\d^0: \quad C^0_f(\fu , \fs_P) \lra Z^0(\fu ,\ca_P).
\tag{4.4}
\end{equation*}
It is obvious from eqs.(4.1) that $\{\psi_\a^{-1}\}\in C^0_f(\fu , \fs_P)$
and
$\{h_\a\psi_\a^{-1} \}\in C^0_f(\fu , \fs_P)$ with $\{h_\a\}\in C^0(\fu , \G_P)$
define the same solution $A=\{A^{(\a )}\}\in Z^0(\fu ,\ca_P)$  of eqs.(1.1a),
which leads to the bijection
\begin{equation*}
Z^0(\fu ,\ca_P)\simeq C^0(\fu , \G_P)\backslash  C^0_f(\fu , \fs_P).
\tag{4.5}
\end{equation*}
The gauge group $\cg_P$ acts on $\psi^{-1} =\{\psi_\a^{-1}\}$ by the right
multiplication: $\psi^{-1}\mapsto \psi^{-1}g =\{\psi_\a^{-1}g_\a\}$,
$g=\{g_\a\}\in Z^0(\fu ,\fs_P)\equiv H^0(X ,\fs_P)$,
and the definition (3.4) of the moduli space of flat connections can be
reformulated in terms of the set $C_f^0(\fu ,\fs_P)$ by using the bijection
(4.5),
\begin{equation*}
\cm_P\simeq C^0(\fu , \G_P)\backslash  C^0_f(\fu , \fs_P)/Z^0(\fu ,\fs_P).
\end{equation*}

\subsection{Functional matrix equations}

Having a solution $\psi =\{\psi_\a\}$ of eqs.(4.2), one can introduce
matrices $\hat f_{\a\b}=\psi^{-1}_\a f_{\a\b}\psi_\b$ defined on
$U_\a\cap U_\b\ne\vn$. Using eqs.(4.2), it is not difficult to show that
$d\hat f_{\a\b}=0$, and therefore $\hat ff^{-1}=
\{\hat f_{\a\b}f_{\a\b}^{-1}\}\in C^1(\fu ,
\G_P)$. Moreover, it is easy to see that $\hat f_{\a\b}$'s satisfy
equations $\hat f_{\a\b}\hat f_{\b\g}\hat f_{\g\a}=1$
and therefore  $\{\hat f_{\a\b}\}$ can be considered
as transition functions of a locally constant bundle $\hat P\to X$
which is topologically equivalent to the bundle $P$ with the transition
functions $\{f_{\a\b}\}$. Thus, we obtain  a map of the data
$(f, A)$ into the data $(f,\hat f)$, i.e. the connection $A$ is
encoded into the transition functions $\hat f$ of the locally constant
bundle $\hat P$.

Let us now consider ``free'' 1-cochains $\hat f=\{\hat f_{\a\b}\}$
such that $\hat f f^{-1}\in C^1(\fu ,
\G_P)$ and {\it functional} equations
\begin{equation*}
\hat f_{\a\b}\hat f_{\b\g}\hat f_{\g\a}=1
\tag{4.6a}
\end{equation*}
on $U_\a\cap U_\b\cap U_\g\ne\vn$. The space of solutions to
eqs.(4.6a) is isomorphic to the space $Z^1(\fu , \G_P)$ of
1-cocycles. Denote by $Z^1_f(\fu , \G_P)$ the subset of those
solutions $\hat f$ to eqs.(4.6a) for which there exists a splitting
\begin{equation*}
\hat f_{\a\b}= \psi^{-1}_\a f_{\a\b}\psi_\b
\tag{4.6b}
\end{equation*}
with some $\psi =\{\psi_\a\}\in C^0(\fu ,\fs_P)$. Using equations
$d\hat f_{\a\b}=0$, one can easily show that these $\{\psi_\a\}$ will
satisfy eqs.(4.2), i.e. $\psi =\{\psi_\a\}\in C_f^0(\fu ,\fs_P)$.
Then, by introducing $A^{(\a)}=\d^0\psi_\a =-(d\psi_\a
)\psi_\a^{-1}$, we obtain a flat connection $A=\{A^{(\a )}\}$ on $P$.
Thus, for constructing flat connections on $P$, one can solve either
{\it differential} equations (4.2) on $\{\psi_\a\}\in C^0(\fu
,\fs_P)$ or {\it functional} equations (4.6) on $\{\hat f_{\a\b}\}$ such
that $\hat f f^{-1}\in C^1(\fu ,\G_P)$.

\subsection{Solution spaces $\ \lra\ $ moduli spaces}

In Sections 4.1, 4.2 we showed that for the fixed transition functions
$f=\{f_{\a\b}\}$ of the locally constant bundle $P$ there is a
correspondence
\begin{equation*}
Z^0(\fu ,\ca_P)\ni (f, A)\quad \leftrightarrow\quad
(f, \hat f)\in Z^1_f(\fu ,\G_P),
\tag{4.7}
\end{equation*}
where $Z^0(\fu ,\ca_P)$  is the solution space of differential equations
(1.1a), and $Z^1_f(\fu ,\G_P)$ is the solution
space of functional equations (4.6). These solution spaces are not bijective.
Namely, to gauge equivalent flat connections $A$ and $\Ad_{g^{-1}}A=
g^{-1}Ag+g^{-1}dg$ with $g=\{g_\a\}\in Z^0(\fu ,\fs_P)$ there corresponds
the same cocycle $\hat f=\{\hat f_{\a\b}\}\in Z^1_f(\fu ,\G_P)$, since
$\hat f^g_{\a\b}:=(\psi^g_\a )^{-1}f_{\a\b}\psi^g_\b =
\psi_\a^{-1}g_\a     f_{\a\b}g_\b^{-1}
\psi_\b = \psi_\a^{-1} f_{\a\b}\psi_\b =\hat f_{\a\b}.$ At the same time,
the cocycles $\hat f$ and $\hat f^h=\{h_\a\hat f_{\a\b}h^{-1}_\b\}$ with
$h=\{h_\a\}\in C^0(\fu ,\G_P)$ correspond to the same flat connection $A$.

The correspondence (4.7) will become the bijection (3.11) if we
consider the space of orbits of the group $Z^0(\fu ,\fs_P)=\cg_P$ (the
gauge group of the model (1.1a)) in the space $Z^0(\fu ,\ca_P)$
and consider the space of orbits of the group $C^0(\fu ,\G_P)$ (the gauge
group of the model (4.6)) in the space $Z^1_f(\fu ,\G_P)$. In other
words, we should consider the pairs $(f,[A])$ and $(f,[\hat f])$,
where $[A]$ is the gauge equivalence class of $A$, and $[\hat f]$ is the
equivalence class of the 1-cocycle $\hat f$. In Sect.3 this correspondence
was described in more abstract terms.

We obtain the following correspondence between de Rham and \v{C}ech
description of flat connections:
\begin{equation*}
\begin{array}{@{}c@{}}
C^0_f(\fu ,\fs_P)
\\[1ex]
\d^0\swarrow\quad\quad\searrow r
\\[1ex]
Z^0(\fu ,\ca_P)\quad \quad Z^1_f(\fu ,\G_P)
\\[1ex]
\pi\downarrow\qquad \qquad \downarrow p
\\[1ex]
\cm_P=\frac{Z^0(\fu ,\ca_P)}{ Z^0(\fu ,\fs_P)}\longleftrightarrow
\frac{Z^1_f(\fu ,\G_P)}{C^0(\fu ,\G_P)}\qquad. \quad
\end{array}
\tag{4.8}
\end{equation*}
Here a map
\begin{equation*}
r:\quad C^0_f(\fu ,\fs_P)\lra   Z^1_f(\fu ,\G_P)
\tag{4.9a}
\end{equation*}
is given by the formula $r(\{\psi_\a\})=\{\psi_\a^{-1}f_{\a\b}
\psi_\b\}=\hat f$,  $\pi$ and $p$ are projections
\begin{gather*}
\pi :\quad Z^0(\fu ,\ca_P)\lra {Z^0(\fu ,\ca_P)}/{ Z^0(\fu ,\fs_P)},
\tag{4.9b}
\\
p:\quad  Z^1_f(\fu ,\G_P)\lra  {C^0(\fu ,\G_P)}\backslash{Z^1_f(\fu ,\G_P)}                ,
\tag{4.9c}
\end{gather*}
and a map $\d^0$ is the projection (4.4). Recall that the group
$Z^0(\fu ,\fs_P)$ acts on the spaces   $C^0_f(\fu ,\fs_P)\ni
\{\psi_\a^{-1}\}$ and $Z^0(\fu ,\ca_P)$ on the right, and the group
$C^0(\fu ,\G_P)$ acts on the spaces $C^0_f(\fu ,\fs_P)\ni
\{\psi_\a^{-1}\}$ and $Z^1_f(\fu ,\G_P)$ on the left.

Using a bijection of the moduli spaces of solutions to eqs.(4.2), (1.1a)
and (4.6), we identify the points $\pi\circ\d^0(\psi ), \pi (A), p(\hat f)$
and $p\circ r(\psi )$, where $\psi\in C^0_f(\fu ,\fs_P), A\in Z^0(\fu ,
\ca_P), \hat f\in  Z^1_f(\fu ,\G_P)$. Then we have
\begin{gather*}
\pi\circ\d^0(\psi )=\pi (A)=p(\hat f)=p\circ  r(\psi )   \Longleftrightarrow
\\
\hat f=s\circ\pi (A)=r\circ\vp(A),\quad
A=\s\circ p(\hat f)=\d^0\circ \eta(\hat f),\quad
\psi =\vp (A)=\eta(\hat f),
\tag{4.10}
\end{gather*}
where $\vp$, $\eta$, $\sigma$ and $s$ are some (local) sections of
the fibrations (4.4), (4.9a), (4.9b) and (4.9c), respectively. Let us
emphasize that just the ambiguity of the choice  of sections of the
fibrations (4.4) and (4.9) leads to the ambiguity of finding $\hat f$
for a given $A$ and $A$ for a given $\hat f$. As usual, one can
remove this ambiguity by choosing some {\it special}
sections $\vp$, $\eta$, $\sigma$ and $s$ (gauge fixing).

\section{Symmetry transformations and deformations of bundles}

In Sect.4 we have described the transformation $r\circ \vp :
(f,\psi d\psi^{-1})\mapsto (f,\hat f)$. If we now define a map
$T_1:\hat f\mapsto\tilde f$ preserving the set $Z^1_f(\fu ,\G_P)$,
then after using the maps $\sigma\circ p$ or $\d^0\circ\eta$, we obtain
a new flat connection $\tilde A$. Below we describe this method of
constructing flat connections on $P$ based on deformations
$T_1:\hat f\mapsto\tilde f$ of locally constant bundles. Namely,
starting from the \v{C}ech approach, in
Sect.5.1 we describe groups acting on the space of transition functions
of a locally constant bundle. Then, using solutions of the functional
matrix equations, we proceed in Sect.5.2 to the
de Rham description and define dressing transformations acting on the
space of solutions to the field equations
of Chern-Simons and topological BF theories. Finally,
in Sect.5.3 we discuss a special class of solutions of the functional
matrix equations describing deformations of flat bundles.

\subsection{Cohomological groups acting on solution spaces}

{} To begin with, we shall consider  groups acting on the set
$C^1(\fu ,\G_P)$ of 1-cochains. One of such groups - the group
$C^0(\fu ,\G_P)$ - and its left action $T_0$ on $C^1(\fu ,\G_P)$, $Z^1(\fu
,\G_P)$ and $Z^1_f(\fu ,\G_P)$ have been considered in Sections 2.2 and
4.3. The action $T_0$ of the group $C^0(\fu ,\G_P)$ gives trivial
(gauge) transformations of the transition functions $\hat f=\{\hat
f_{\a\b}\}$, and the space of orbits of the group $C^0(\fu ,\G_P)$ in
$Z^1_f(\fu ,\G_P)$ is the moduli space $\cm_P\simeq C^0(\fu
,\G_P)\backslash Z^1_f(\fu ,\G_P)=\Ker \rho\subset H^1(X,
\G_P)$. In other words, the group $C^0(\fu ,\G_P)$ is the stability
subgroup of the point $[\hat f]\in\cm_P$.

To obtain a nontrivial map of the moduli space $\cm_P$ onto itself, we
consider the following action of the group $C^1(\fu ,\G_P)$ on itself:
\begin{equation*}
T_1(h,\ .\ ):\ \chi\mapsto T_1(h,\chi ),
\quad T_1(h,\chi )_{\a\b}:=
h_{\a\b}\chi_{\a\b}(h^{(\a )}_{\b\a})^{-1},
\tag{5.1}
\end{equation*}
where $h,\chi\in C^1(\fu ,\G_P)$. Recall that $h_{\b\a}^{(\a )}\ne
h^{-1}_{\a\b}$ if $h\not\in Z^1(\fu , \G_P)$. From the definition
(5.1) it is easy to see that
$T_1(g, T_1(h,\chi ))=T_1(gh,\chi )$ for $g,h,\chi\in C^1(\fu ,\G_P)$.
Moreover, starting from the element $\chi\in C^1(\fu ,\G_P)$,
one can obtain any other element of $C^1(\fu ,\G_P)$, i.e. the action
$T_1$ is transitive.

We are interested in subgroups of the group $C^1(\fu ,\G_P)$ preserving
the set $Z^1_f(\fu ,\G_P)\subset Z^1(\fu ,\G_P)\subset C^1(\fu ,\G_P)$.
To describe them, we fix $\hat f f^{-1}=\{\hat f_{\a\b}f_{\a\b}^{-1}\}=
\{\psi^{-1}_\a f_{\a\b}\psi_\b f_{\a\b}^{-1}\}\in Z^1_f(\fu ,\G_P)$ and
consider an orbit of the point $\chi =\hat ff^{-1}$
under the action $T_1$ of the group  $C^1(\fu ,\G_P)$. We want to find
an intersection of this orbit with the set $Z^1_f(\fu ,\G_P)$, i.e. the set
of all $h\in C^1(\fu ,\G_P)$ such that
\begin{itemize}
\item[(i)] $\tilde ff^{-1}=T_1(h,\hat ff^{-1})\in Z^1(\fu ,\G_P)$, i.e. the
transformation $T_1(h,\ .\ ): \hat f\mapsto\tilde f=
h_{\a\b}\hat f_{\a\b}h_{\b\a}^{-1}$ preserves eqs.(4.6a),
\item[(ii)] $\tilde ff^{-1}=T_1(h,\hat ff^{-1})\in Z^1_f(\fu ,\G_P)\subset
Z^1(\fu ,\G_P) $, i.e. the transformation $T_1(h,\ .\ ): \hat f\mapsto
\tilde f=h_{\a\b}\hat f_{\a\b}h_{\b\a}^{-1}$ preserves not only eqs.(4.6a),
but also eqs.(4.6b).
\end{itemize}

The condition (i) means that the transformation $T_1(h,\ .\ )$ maps
$Z^1(\fu ,\G_P)$ onto itself, which is realized, of course, not for any
$h\in C^1(\fu ,\G_P)$. This condition imposes  severe constraints
on $h$ which are equivalent to the following nonlinear functional equations:
\begin{equation*}
h_{\a\b }\ \hat f_{\a\b }\ h_{\b\a }^{-1}\ h_{\b\g }\
\hat f_{\b\g }\ h_{\g\b }^{-1}\ h_{\g\a }\
\hat f_{\g\a }\ h_{\a\g }^{-1}=1
\quad \mbox{on}\quad U_\a\cap U_\b\cap U_\g\ne\vn .
\tag{5.2a}
\end{equation*}
Notice that eqs.(5.2a) are satisfied trivially if $h\in C^0(\fu ,\G_P)$,
since in this case $h_{\a\b}=h_{\a |\b}:=h_{\a |U_\a\cap U_\b}$,
$h_{\b\a}=h_{\b |\a}:=h_{\b |U_\a\cap U_\b}$.

By introducing $\chi_{\a\b}=\tilde f_{\a\b}f^{-1}_{\a\b}$ and using
formulae (2.5), we can rewrite eqs.(5.2a) in the form
\begin{equation*}
\chi_{\a\b}^{(\a )}\chi_{\b\g}^{(\a )}\chi_{\g\a}^{(\a )}=1
\quad \mbox{on}\quad U_\a\cap U_\b\cap U_\g\ne\vn .
\tag{5.2b}
\end{equation*}
{}Formulae (5.2) mean that $\chi =\{\chi_{\a\b}\}=\{\chi_{\a\b}^{(\a )}\}$
is a 1-cocycle, $\chi\in Z^1(\fu , \G_P)$.

The condition (ii) means that for the 1-cocycle $\tilde f$
one can find a splitting like (4.6b), i.e. there exists a 0-cochain
$\tilde\psi =\{\tilde\psi_\a\}\in C^0 (\fu ,\fs_P)$ such that
\begin{equation*}
\tilde f_{\a\b}=\tilde\psi^{-1}_\a f_{\a\b} \tilde\psi_\b  .
\tag{5.3}
\end{equation*}
This imposes additional restrictions on $h\in C^1(\fu ,\G_P)$, and
one of their possible resolvings is the following. Recall that for
the sheaf $\frs_{\hat P}$ of smooth sections of the bundle $\ad \hat
P$ we have $H^1(X,\frs_{\hat P})=0$~\cite{GrH}. Therefore for {\it
small} deformations $\hat f\mapsto\tilde f$ of the cocycle $\hat f$
there always exists $\tilde\psi =\{\tilde\psi_\a\}\in C^0(\fu
,\fs_P)$ such that (5.3) is satisfied. As such, we shall consider
elements $h\in C^1(\fu ,\G_P)$ that are close to the identity and
satisfying eqs.(5.2). In other words, we consider a {\it local group}
(a neighbourhood of the identity in $C^1(\fu ,\G_P)$) and take only
those its elements $h$ which satisfy the conditions (5.2). If,
in addition, we change the ``initial'' point $\hat f\in Z^1_f(\fu
,\G_P)$ for the action $T_1$, then in the described way we shall
obtain all elements of the group of bijections of the set $Z^1_f(\fu
,\G_P)$.

\subsection{Dressing transformations}

In Sect.4 we have described maps $A\mapsto\hat f$ and $\hat f\mapsto
A$, where $A=\psi d\psi^{-1}\in Z^0(\fu , \ca_P)$, $\hat f =\{\hat
f_{\a\b}\}= \{\psi^{-1}_\a f_{\a\b}\psi_\b\}\in Z^1_f(\fu ,\G_P)$,
$\psi=\{\psi_\a\} \in C^0_f(\fu ,\fs_P)$. In Sect.5.1 we have
described the map $T_1(h,\ . \ ):  \hat f\mapsto\tilde
f=\{h_{\a\b}\hat f_{\a\b}h^{-1}_{\b\a}\},$ where $\tilde f\in
Z^1_f(\fu ,\G_P)$ if elements $h=\{h_{\a\b}\}\in C^1(\fu , \G_P)$
satisfy some conditions. Then, having found $\tilde\psi =
\{\tilde\psi_\a\}$ by formula (5.3), we introduce
\begin{equation*}
\tilde A^{(\a
)}:=-(d\tilde\psi_\a) \tilde\psi_\a^{-1}.  \tag{5.4}
\end{equation*}
{}From the
equations $d\tilde f_{\a\b}=0$ and eqs.(5.3) it follows that $\tilde
\psi_\a$'s satisfy eqs.(4.2), and $\tilde A^{(\a)}$'s satisfy
eqs.(4.1b), i.e.
\begin{equation*}
\tilde A^{(\a)}=f_{\a\b}\tilde A^{(\b)}
f_{\a\b}^{-1}, \tag{5.5a}
\end{equation*}
\begin{equation*}
(d\tilde\psi_\a )\tilde\psi_\a^{-1}=
f_{\a\b}(d\tilde\psi_\b ) \tilde\psi_\b^{-1}f_{\a\b}^{-1} \tag{5.5b}
\end{equation*}
on $U_\a\cap U_\b\ne\vn$. Thus, if we take a flat connection
$A=\{A^{(\a)}\}$ and carry out the sequence of transformations
\begin{equation*}
(f,A)\stackrel{r\circ\vp}\mapsto (f,\hat f)\stackrel{T_1}\mapsto
(f,\tilde f)\stackrel{\d^0\circ \eta}\longmapsto (f,\tilde A),
\tag{5.6}
\end{equation*}
we obtain a new flat connection $\tilde A=\{\tilde
A^{(\a )}\}$. Notice that by virtue of commutativity of the diagram
(4.8), one can consider the map $s\circ\pi$ instead of $r\circ\vp$
and the map $\sigma\circ p$ instead of $\d^0\circ\eta$.

It is not difficult to verify that
\begin{equation*}
\tilde A=\{\tilde A^{(\a )}\}=\{\phi_\a A^{(\a )}\phi^{-1}_\a  +
\phi_\a d\phi^{-1}_\a\}= \phi A\phi^{-1}  + \phi d\phi^{-1},
\tag{5.7a}
\end{equation*}
where
\begin{equation*}
\phi :=\tilde\psi\psi^{-1}= \{\tilde\psi_\a\psi^{-1}_\a\}=\{\phi_\a\}\in
C^0(\fu ,\fs_P).
\tag{5.7b}
\end{equation*}
{}Formally, (5.7) looks like a gauge transformation.  But actually the
transformation $\Ad_\phi : A\mapsto\tilde A$, defined by (5.7), consists of
the sequence (5.6) of transformations and is not a gauge transformation,
 since $\phi_\a\ne f_{\a\b}\phi_\b f_{\a\b}^{-1}$ on $U_\a\cap
U_\b\ne\vn$. Recall that for gauge transformations $\Ad_g:  A\mapsto
A^g=gAg^{-1} +gdg^{-1}$ one has $g_\a =f_{\a\b}g_\b f_{\a\b}^{-1}$, i.e.
$g=\{g_\a\}$ is a {\it global} section of the bundle $\Int P$, and
$\phi=\{\phi_\a\}$ is a collection of {\it local} sections $\phi_\a:
U_\a\to G$  of the bundle $\Int P$ which are constructed by the
algorithm described above.

{}From formulae (3.13), (4.3) and (5.7b) it follows that the dressing
transformation $\Ad_{\phi}$ acts on any solution $B$ of eqs.(1.1b) by the
formula
\begin{equation*}
\Ad_\phi :\ B\mapsto\ \tilde B=\{\tilde B^{(\a )}\}:=\{\phi_\a B^{(\a )}
\phi_\a^{-1}\}=\phi B\phi^{-1},
\tag{5.7c}
\end{equation*}
where $\phi =\{\phi_\a\}$ is defined in (5.7b). The field $\tilde B$
is a new solution of eqs.(1.1b). As is shown above, the transformation
$\Ad_\phi$ is not a gauge transformation.

Let us emphasize that  $\phi=\{\phi_\a\}$ depends on $h\in C^1(\fu
,\G_P)$ defining the transformation $T_1(h,\ .\ ): \hat
f\mapsto\tilde f$, and taking this into account we shall write
$\phi\equiv\phi (h)=\{\phi_\a(h)\}$. The transformations (5.7) will
be called the {\it dressing transformations}.  In this terminology we
follow the papers~\cite{ZS,STS,BB}, where the analogous
transformations were used for constructing solutions of integrable
equations. It is not difficult to show that these transformations
form a group.

Notice that the maps (5.6) are connected with maps between
the bundles $(P,f)$, $(\hat P,\hat f)$ and $(\tilde P, \tilde f)$.
All these bundles are diffeomorphic but not isomorphic as locally
constant bundles. A diffeomorphism of $P$ onto $\hat P$ is defined
by a 0-cochain $\psi =\{\psi_\a\}\in C^0_f(\fu ,\fs_P)$, and a
diffeomorphism of $P$ onto $\tilde P$ is defined by $\tilde\psi =\{
\tilde\psi_\a\}\in C^0_f(\fu ,\fs_P)$. Moreover, the bundles $\hat P$
and $\tilde P$ become isomorphic as locally constant bundles after the
restriction to $U_\a\cap U_\b\ne\vn$: $\hat P|_{U_\a\cap U_\b}\simeq
\tilde P|_{U_\a\cap U_\b}$, but these isomorphisms are different for
${U_\a\cap U_\b}$ with different $\a ,\b\in I$. In other words,
$h_{\a\b}:  \hat P|_{U_\a\cap U_\b}\to \tilde P|_{U_\a\cap U_\b}$ define
{\it local} isomorphisms that do not extend up to the isomorphism
of $\hat P$ and $\tilde P$ as locally constant bundles over the whole $X$.

\subsection{A special cohomological symmetry group}

Matrices $h=\{h_{\a\b}\}$, defining the dressing transformations
\begin{gather*}
\Ad_{\phi (h)}:\ A\mapsto A^h:= \phi (h) A \phi (h)^{-1} +\phi (h)d
\phi (h)^{-1},
\tag{5.8a}
\\
\Ad_{\phi (h)}:\ B\mapsto B^h:= \phi (h) B \phi (h)^{-1},
\tag{5.8b}
\end{gather*}
must satisfy the nonlinear functional equations (5.2) which are not
so easy to solve. However, there exists an important class of
solutions to these equations, which can be described explicitly. The
merit of these solutions is the fact that they do not depend on the
choice of cocycles $f$ and $\hat f$.

Notice that eqs.(5.2) with all indices written down have the form
\begin{equation*}
h_{\a\b |\g}\ \hat f_{\a\b |\g}\ h_{\b\a |\g}^{-1}\ h_{\b\g |\a}\
\hat f_{\b\g |\a}\ h_{\g\b |\a}^{-1}\ h_{\g\a |\b}\
\hat f_{\g\a |\b}\ h_{\a\g |\b}^{-1}=1,
\tag{5.9}
\end{equation*}
where $h_{\a\b |\g}$ means the restriction of $h_{\a\b}$ defined on
$U_\a\cap U_\b$ to an open set $U_\a\cap U_\b\cap U_\g$ and analogously
for all other matrices. It is not difficult to verify that a collection
$h=\{h_{\a\b}\}\in C^1(\fu ,\G_P)$ of matrices such that
\begin{equation*}
h_{\a\b |\g}=h_{\a\g |\b}, \quad
h_{\b\a |\g}=h_{\b\g |\a}, \quad
h_{\g\b |\a}=h_{\g\a |\b}
\tag{5.10}
\end{equation*}
satisfies eqs.(5.9) for any choice of 1-cocycles $f=\{f_{\a\b}\}$ and
$\hat f=\{\hat f_{\a\b}\}$.

The constraints (5.10) are not very severe. They simply mean that
sections $h_{\a\b}$ of the sheaf $\G_P$ over $U_\a\cap U_\b\ne\vn$
can be extended to sections of the sheaf $\G_P$ over the open set
\begin{equation*}
\U=\bigcup\limits_{\a ,\b\in I}U_\a\cap U_\b ,
\tag{5.11}
\end{equation*}
where the summation is carried out in all  $\a ,\b\in I$ for which
$U_\a\cap U_\b\ne\vn$. In other words, from (5.10) it follows that
there exists a locally constant section $h_\U: \U\to G$ of the
bundle $\Int P$ such that $h_{\a\b}=h_{\U|U_\a\cap U_\b}$
\cite{GrH}. In this case we can identify $h=\{h_{\a\b}\}=
\{h_{\U|U_\a\cap U_\b}\}$  and $h_\U$. Such $h$ form a subgroup
\begin{equation*}
\bar C^1(\fu ,\G_P):=\{h\in C^1(\fu ,\G_P): h_{\a\b |\g}=h_{\a\g|\b}\
\mbox{on} \ U_\a\cap U_\b\cap U_\g\ne\vn\}
\tag{5.12}
\end{equation*}
of the group $C^1(\fu,\G_P)$.

Now let us consider  $\bar C^1(\fu ,\G_P)$ as a {\it local } group,
i.e. let us take $h$ from the neighbourhood of the identity in
$\bar C^1(\fu ,\G_P)$. For such $h$ there always exists $\tilde\psi\in
C^0(\fu ,\fs_P)$ such that (5.3) is satisfied and one can introduce
$\phi (h)$ by formula (5.7b). Notice that this map $\phi :
\bar C^1(\fu ,\G_P)\to C^0(\fu ,\fs_P)$ is a homomorphism. Using
$\phi (h)$, we can introduce a new flat connection $A^h=\Ad_{\phi (h)}A=
\phi (h)A\phi (h)^{-1}  + \phi (h)d\phi (h)^{-1}$
and a new solution $B^h=\Ad_{\phi (h)}B=\phi (h)B\phi (h)^{-1}$ of
eqs.(1.1b) for any $h\in\bar C^1(\fu ,\G_P)$.

\section{Conclusion}

In this paper, the group-theoretic analysis of the moduli space of
solutions of Chern-Simons and topological BF theories  and nonlocal
symmetries of their equations of motion has been undertaken. We
have formulated
a method of constructing flat connections based on a correspondence
between the non-Abelian de Rham and \v{C}ech cohomologies. The
group of all symmetries of the field equations of Chern-Simons and
topological BF theories
and the special cohomological symmetry group have been described. The
described dressing symmetries can be lifted up to symmetries of
quantum Chern-Simons and topological BF theories. It would be desirable
to describe and analyze representations of these symmetry groups.

\subsection*{Acknowledgements}

This work is supported in part by the grant RFBR-99-01-01076.
A.D.P. is  grateful to the CERN Theory Division for hospitality
during the final stage of the work.


\label{lastpage}

\end{document}